\documentclass[preprint,aps,pre,showpacs,epsfig,amsmath,amssymb]{revtex4}
\usepackage{makecell}
\usepackage{bbm}
\usepackage{mathrsfs}
\usepackage{subfigure}
\usepackage{amsfonts}
\usepackage{color}
\usepackage{graphicx}

\def\newblock{\hskip .11em plus .33em minus .07em}

\newcommand{\be}{\begin{equation}}
\newcommand{\ee}{\end{equation}}
\newcommand{\ba}{\begin{eqnarray}}
\newcommand{\ea}{\end{eqnarray}}

\begin{document}
\title{Information Swimmer: Self-propulsion without Energy Dissipation}
\author{Chen Huang$^{1}$}
\author{Mingnan Ding$^{1}$}
\author{Xiangjun Xing$^{1,2,3}$}
\email{xxing@sjtu.edu.cn}
\address{$^1$ Wilczek Quantum Center, School of Physics and Astronomy, Shanghai Jiao Tong University, Shanghai, 200240 China\\
$^2$ T.D. Lee Institute, Shanghai Jiao Tong University, Shanghai, 200240 China \\
$^3$ Shanghai Research Center for Quantum Sciences, Shanghai 201315 China}
	
	
	
\begin{abstract}

We study an information-based mechanism of self-propulsion in noisy environment.  An {\em information swimmer} maintains directional motion by periodically measuring its velocity and accordingly adjusting its friction coefficient. Assuming that the measurement and adjustment are {\it  reversible} and hence cause no energy dissipation, an information swimmer may move without external energy input.  There is however no violation of the second law of thermodynamics, because the information entropy stored in the memory of swimmer increases monotonically. By optimizing its control parameters, the swimmer can achieve a steady velocity that is comparable to the root-mean-square velocity of an analogous Brownian particle. 
 We also define a swimming efficiency in terms of information entropy production rate, and find that in equilibrium media with white noises, information swimmers are generally less efficient than Brownian particles driven by constant forces. For colored noises with long correlation times, the frequency of measurement can be greatly reduced without affecting the efficiency of information swimmers.  
	
\end{abstract}
	
\maketitle 

{\bf Introduction} \quad {Self-propulsion at micro and nano scales are intensively studied, both as a problem of biological physics~\cite{trepat2018mesoscale,di2010bacterial,moran2017phoretic}, and as a problem of bio-inspired engineering~\cite{moran2017phoretic, ebbens2010pursuit,goldman2014colloquium, bintein2019self,palacci2013living,morozov2017chaos,cordova2008osmotic,lagubeau2011leidenfrost,hanggi1996brownian,astumian2002brownian,wang2013small,mano2005bioelectrochemical,stone1996propulsion}.  } While self-propelling of bacteria is typically achieved via actuation of cellular appendages such as flagella, synthetic self-propellors often move via surface effects~\cite{wang2013small,stone1996propulsion}, or phoretic effects~\cite{moran2017phoretic}, i.e., interaction with gradient of physical quantities.  Another interesting self-propelling mechanism is {\em Brownian motor}~\cite{astumian2002brownian,hanggi1996brownian}, which relies on a delicate interplay between noises and periodic potential.  


Here we explore a novel mechanism of self-propulsion that uses information instead of energy. We imagine a swimmer periodically measures its velocity relative to its environment, and adjust its friction coefficient accordingly.  As a consequence it is able to maintain a steady motion along the chosen direction, with an average velocity comparable with root-mean-square velocity of Brownian motion.   We shall call such a system an {\em information swimmer}, in echo of {\em information engine} which use information to extract energy from a single heat bath.

Perpetual motion with no energy dissipation may widely be perceived as violating the second law of thermodynamics.  The essence of the second law is however not about energy, but about entropy.  In the presence of information acquiring devices, entropy increase is not necessarily accompanied by energy dissipation.  The relation between entropy and information is an intellectually profound question with long and interesting history~\cite{leff2002maxwell}.  Through the works of Maxwell~~\cite{knott1911quote} , Szilard~\cite{szilard1929entropieverminderung}, Landauer~\cite{landauer1961irreversibility}, Penrose~\cite{penrose1979foundations}, and Bennett~\cite{bennett1973logical,bennett1982thermodynamics}, and many more recent studies~\cite{leff2002maxwell,sagawa2019second,parrondo2015thermodynamics},  it has become clear that in the presence of information acquiring agent, the total entropy can be written as 
 \be
S_{\rm tot} = H ({\rm Info}) +  S({\rm Sys}|{\rm Info}), 
\label{entropy-decomp}
\ee
whereas $H ({\rm Info}) $ is the information entropy, and $S({\rm Sys}|{\rm Info})$ is the thermodynamic entropy {\em conditioned on the information acquired}.   Thermodynamic entropy and information entropy can be transformed into each other, much like energy and mass. The total entropy is however dictated by the second law to be non-decreasing.  

There have recently been a large body of researches on design and application of ``information heat engines'', which use information to extract mechanical work, or to push particles to higher free energy states \cite{cao2009thermodynamics,lu2014engineering,vaikuntanathan2011modeling,bergli2014accuracy,abreu2011extracting,park2016optimal,paneru2018lossless,paneru2018optimal,toyabe2010experimental}. Information swimmer is an information engine which serves the distinct purpose of maintaining of directional transport in noisy environment. Of special interests is a design called ``information ratchet''~\cite{serreli2007molecular,sagawa2010generalized}, where one measures the position of an object, and adjust the {\it confining potential} accordingly, which leads to directional transport with no apparent energy dissipation. Unlike information ratchet, the mechanism we study in this work does not require a periodic confinement potential.

Information swimmers may be realized using colloidal particle, polymeric materials, or biological molecules.   An information swimmer consists of at least three components: a sensor (measuring velocity),  a memory (storing information), and a switch (controlling friction coefficient).  Accordingly, the working cycle of the swimmer consists of three basis operations: measurement, information storage, and tuning of friction coefficient.  While in reality these operations are always dissipational, there is no lower bound of dissipation imposed by any fundamental law of physics.  The observation that measurement can in principle be made reversible and hence causes no energy dissipation was first made by Bennett~\cite{bennett1973logical,bennett1982thermodynamics,bennett1987demons}, and played an essential role in the proper resolution of paradox raised by Maxwell's demon.  For a review, see reference ~\cite{leff2002maxwell}. 
Likewise, the operation of information storage (which can be understood as a special form of computation) can also be made reversible, and hence causes no energy dissipation.  (Also pointed out by Bennett is that information erasure is always irreversible and hence cause energy dissipation.  We shall further assume that the memory space is sufficiently large so that there is no need of information erasure.)  The friction coefficient may be tuned by changing the swimmer's volume, shape, or surface structure, which can be realized using structure phase transition of polymeric materials.  A fancier way is to deform the particle using microscopic molecular motors~\cite{mavroidis2003pulses,block1998kinesin,kitamura1999single,burgess2003dynein,yurke2000dna,berg1974dynamic,knoblauch2003atp} or nanorobots~\cite{magdanz2014stimuli,stoychev2012shape,stoychev2013hierarchical}.  Again there is no lower bound of energy cost/dissipation in these processes, and hence we will assume them to be reversible.  Assuming that all these operations are reversible, an information swimmer can maintain directional motion without energy dissipation.   This however does not mean violation of the second law, since the information entropy stored in the swimmer's memory does increase steadily.

Swimming on information may have advantages over existing mechanisms of transport in microscopic noisy environment. It does not need externally imposed potential, which is required by information ratchets and Brownian motors, or proximity to interface, which is required by phoretic swimmers. It may have higher biocompatibility since it causes much less (in principle zero) energy dissipation.

It is interesting to note that life uses information for control of transport, long before human understand information.  For example, in chemotaxis~\cite{berg1993random}, bacteria tune their motion using information (together with energy) on gradient of external chemical stimulus (either attractant or repellent).  In swarming, birds and insects adjust their fly according to their distances to neighbors~\cite{camazineself,blum2008swarm}.   In marine navigation, sailors control the directions of rudders and sails~\cite{jobson2005sailing} to make boat turn (tacking) and move (zigzagging) along arbitrary direction relative to the wind.  Sailing is in fact an almost ideal realization of self-propulsion using information only, because the energy cost of turning sails and rudders is negligible comparing with that needed to drive a boat. Further studies may reveal many other information-feedback mechanisms for motion in biological and technological fields.

\begin{figure*}[t!]
	\centering
	\subfigure[]{\includegraphics[height=2.5in]{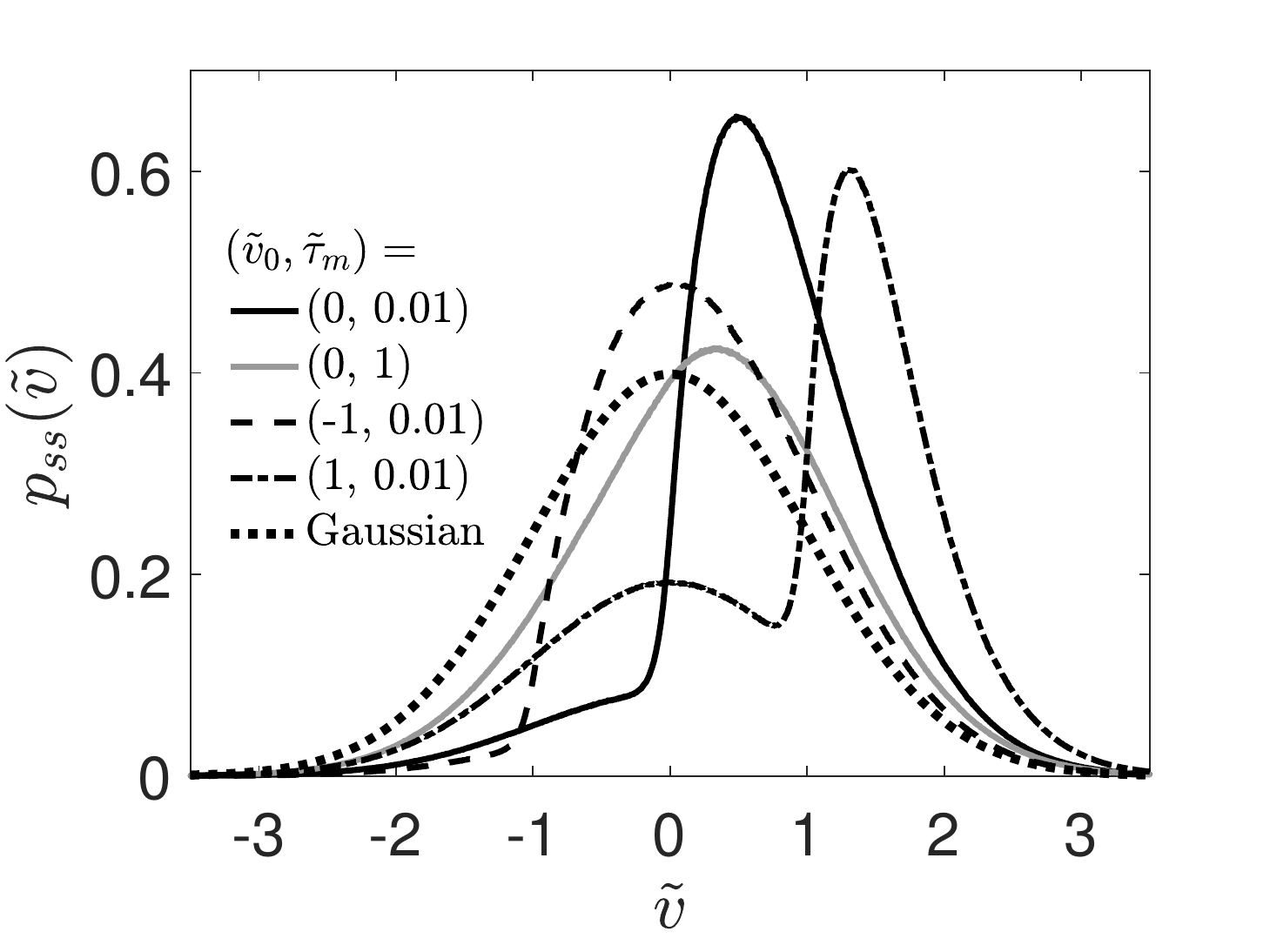}\label{p_ss}}
	\hspace{-6mm}
	\subfigure[]{\includegraphics[height=2.5in]{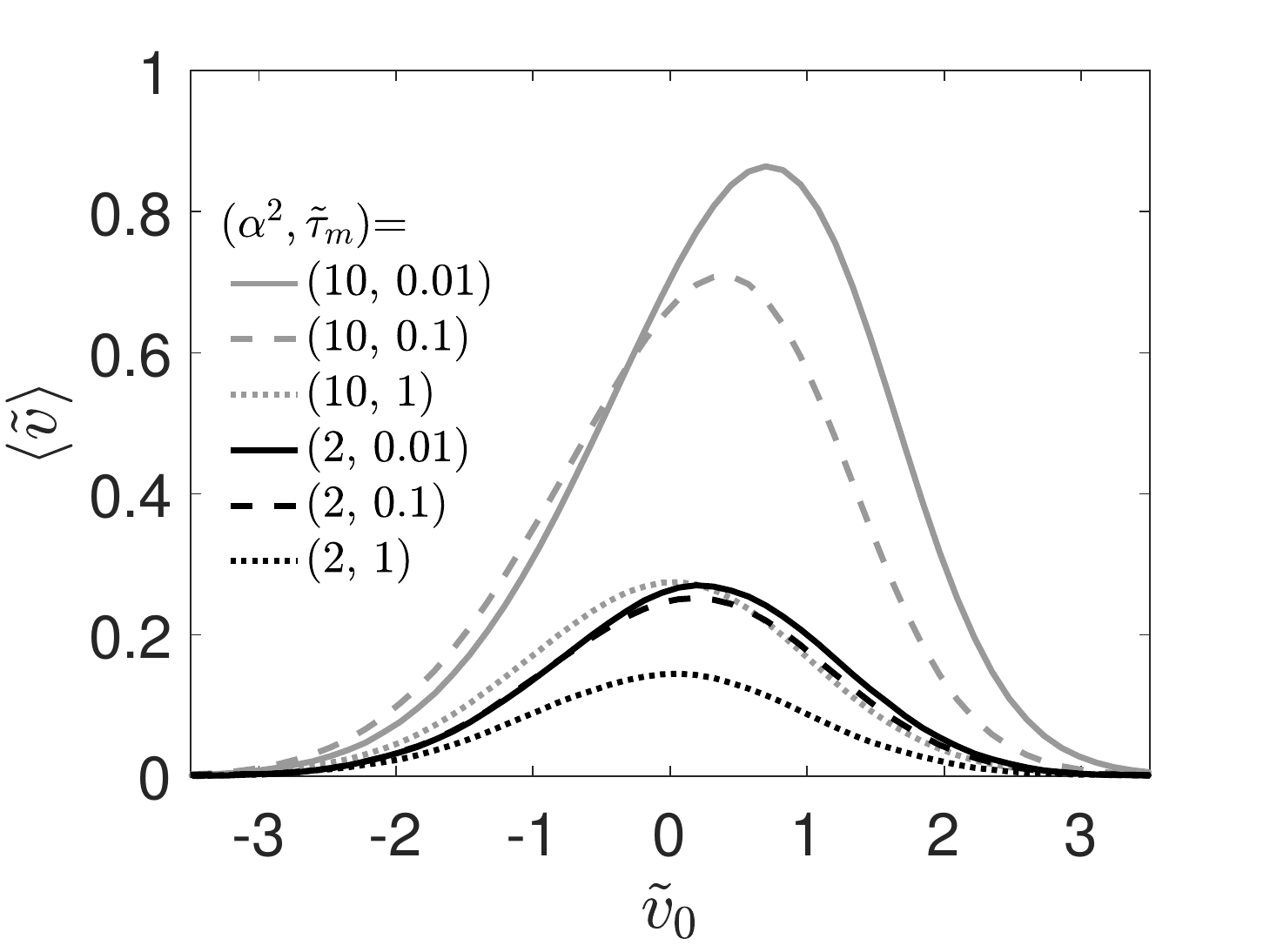}
	\label{v_eff_v_ref}}
	\hspace{-6mm}
	\subfigure[]{\includegraphics[height=3.05in]{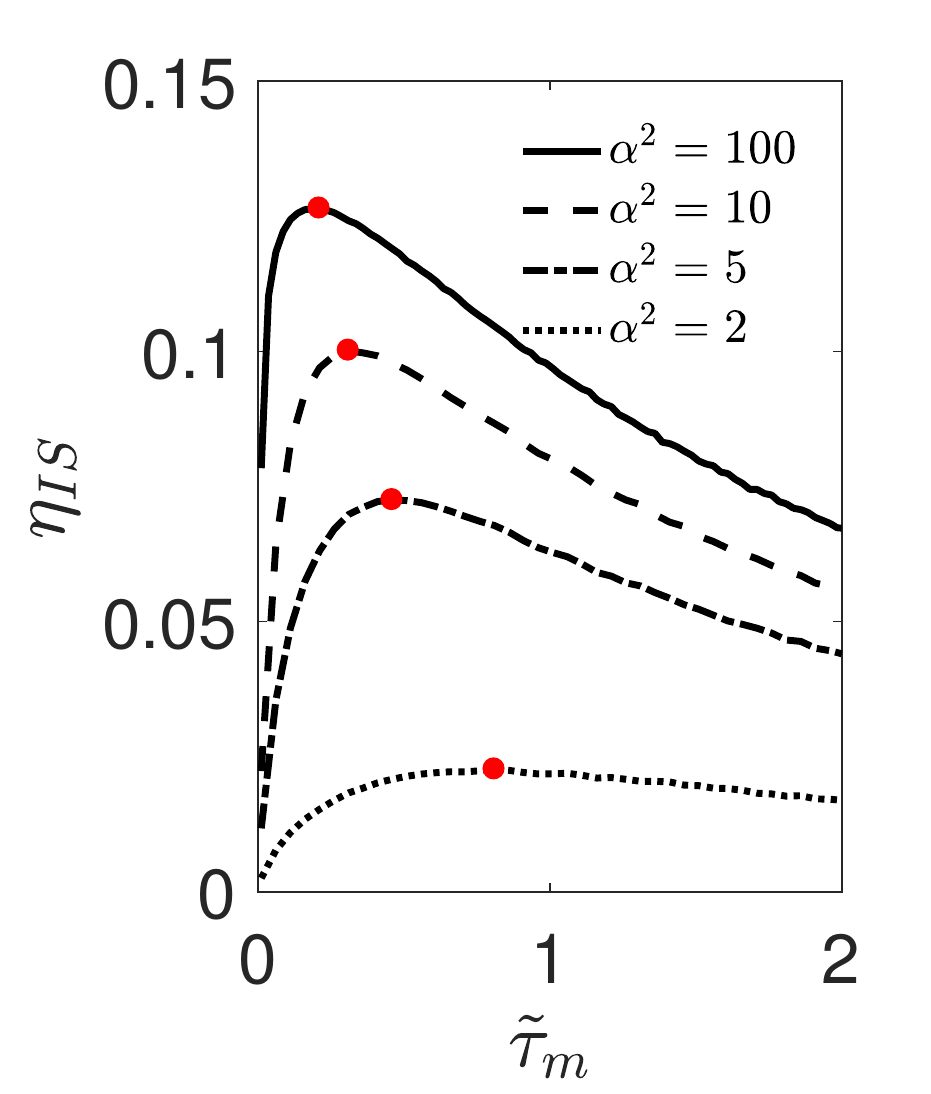}\label{swimmingEfficiency}}
\vspace{-3mm}
	\caption{Information swimming under white noise. (a) Velocity distribution in steady regime at $\alpha^2 = 10$. The dotted line is the Gaussian distribution corresponding to thermal equilibrium. (b) The steady state average velocity $\langle \tilde v \rangle$ as a function of $\tilde{v}_{0}$ at $10$ (gray curves) and $\alpha^2 = 2$ (black curves). Different line types correspond to the value of $\tilde{\tau}_m=0.01$ (solid line), 0.1 (dashed line) and 1 (dotted line). (c) Entropic efficiency $\eta_{IS}$ as the function of $\tilde{\tau}_m$.}
\end{figure*}

\vspace{2mm}

{\bf Model and Simulation Methods} \quad  To reduce the complexity of details, we consider the one-dimensional case.  The physics of higher dimensional cases is essentially the same.  The  swimmer has a baseline friction coefficient  $\gamma$.  After every time interval $\tau_m$, the swimmer measures its velocity and compares with a threshold velocity $v_0$.   The friction coefficient is set to $\gamma$ if $v > v_0$ (the velocity is ``favorable'') and to a large value $ \gamma' = \alpha^2 \gamma > \gamma$ if $v<v_0$ (the velocity is ``unfavorable'').  The results of measurement are recorded in its internal memory space.  

The dynamics of the swimmer can be modeled using piecewise linear Langevin dynamics.  Assuming that the noises acting on the swimmer is Gaussian and white, during the time interval $ n \tau_m < t <(n+1) \tau_m$, the velocity of the  swimmer obeys the following equation (with $k_B=1$):
\ba 
\label{1d}
m\dot {v}(t) =
\left\{\begin{array}{ll}
- \alpha^2 \gamma \, v(t) + \alpha \sqrt {2 \gamma T} \, \zeta(t), 
\quad\quad & v(n \tau_m) \leq v_0,
\vspace{2mm}\\
 - \gamma \, v(t) + \sqrt {2\gamma T} \, \zeta(t) , 
& v (n \tau_m) > v_0 . 
\end{array}
\right.
\label{LE_original}
\ea 
where  $\zeta(t)$ is the normalized Gaussian white noise with statistical properties:
\be
\langle \zeta(t)\rangle= 0,\quad
 \langle\zeta(t)\zeta(t')\rangle= \delta(t-t').
 \label{noise-variance-1}
\ee 
The coefficients of the noise terms in Eq.~(\ref{1d}) are chosen such that {\em the Einstein relation}, i.e., the second Fluctuation-Dissipation relation, is satisfied separately for $v(t) < v_0$ and $v(t) > v_0$.   This relation is a reflection of the equilibrium nature of the ambient fluid, and remains valid independent of the swimmer velocity~\footnote{Of course, we are assuming that the swimmer has negligible influence on the statistical properties of the fluid. }.  If we set $\alpha = 1$,  Eq.~(\ref{1d}) describes a normal Brownian particle, whose velocity distribution converges to a Maxwell distribution with average kinetic energy $T/2$, as required by equilibrium statistical mechanics. We note that the equations of motion are similar to a recently proposed \textit{guided active particle} model~\cite{garcia2019guided}. The main difference from it is that we focus on the information obtained during periodic measurement. This information metric is important to define efficiency in terms of information entropy rate, and to discuss the advantage of colored noise over white noise in the remaining part of the paper.

It is convenient to rescale all variables to obtain a dimensionless theory.  The time scale is just the relaxation time for Brownian motion $\tau = m/\gamma$, whereas the  velocity scale is the typical thermal velocity $v_T = \sqrt{T/m}$.  The dimensionless variables are defined as
\ba
 \tilde{t} \equiv  \frac{t}{\tau}
  , \quad
\tilde{v}(\tilde t)  \equiv  \frac{v(t)}{v_T} , 
\quad
\tilde{\zeta} (\tilde t)\equiv \sqrt{\frac{2 m}{\gamma}} \zeta(t) . 
\label{rescaling-1}
\ea 
Further defining two control parameters $\tilde{v}_0 \equiv {v}_0/v_T$ and $ \tilde \tau_m =  \tau_m/ \tau$, Eqs.~(\ref{1d}) and (\ref{noise-variance-1}) becomes 
\ba
\frac{d \tilde{v}}{d \tilde{t}}  &=&
\left\{\begin{array}{ll}
  - \alpha^2 \tilde{v}(\tilde{t}) + \alpha \, \tilde{\zeta}(\tilde t) , \quad \quad
&  \tilde v (n\tilde \tau_m) \leq \tilde v_0 
\vspace{2mm}\\
 - \tilde{v}(\tilde{t}) + \tilde{\zeta}(\tilde t) , 
& \tilde v (n \tilde \tau_m) > \tilde v_0 ;
\label{LE} 
\end{array}
\right.
 \\
 \langle  \tilde \zeta(\tilde t) \rangle &=& 0,\quad
 \langle \tilde \zeta(\tilde t) \tilde \zeta(\tilde t') \rangle 
 = 2 \, \delta(\tilde t- \tilde t').
 \nonumber
\ea

We use numerical scheme to discretize Eq.~(\ref{LE}) and simulate the trajectory.  The technical details are discussed in Appendix A.  There are three dimensionless control parameters in our model: $\tilde \tau_m$, $\alpha^2$ and $\tilde v_0$.    We compute the steady state velocity distribution and its average via  $\langle \tilde v \rangle\equiv\int_{-\infty}^\infty \tilde{v}p_{ss}(\tilde{v})d\tilde{v}$, and analyze their behaviors as all three parameters are varied.   

In long time, the swimmer reaches a steady moving state. By simple dimensional argument, we expect that the average velocity scales with the thermal velocity $v_T= \sqrt{T/m}$, if we choose $\tilde \tau_m$ smaller than unity, and $\alpha^2$ larger than unity.   For a swimmer with a micron radius moving in a fluid with viscosity comparable to water, we estimate that the time scale $\tau \sim 8 \mu s$ and the steady velocity  $\langle v \rangle \sim 100 \mu m/s$.

\vspace{2mm}
{\bf Velocity distribution} \quad We consider the case $\alpha^2  = 10$, which means that the friction coefficient becomes ten times larger if the velocity is unfavorable. First we set the threshold velocity $\tilde v_0 = 0$, and plot the velocity distributions for two different periods of measurement $\tilde \tau_m=0.01$ and 1.  As shown in Fig.~\ref{p_ss}, for $\tilde{\tau}_m = 0.01$ (one measurement every $\tau/100$ seconds), the velocity distribution has an abrupt change of slope near $\tilde{v} = 0$, and a high peak to the right. The probability density is severely suppressed for $\tilde{v} <0$. The overall shape is drastically different from the equilibrium Gaussian distribution.  The average velocity of the swimmer is approximately $0.7 v_T$, as one can see from Fig.~\ref{v_eff_v_ref}.  For $\tilde{\tau}_m = 1$ (one measurement every $\tau$ seconds), the velocity distribution has a much more regular shape, even though the difference with equilibrium distribution is clearly noticeable.  The average velocity is approximately $0.25 v_T$, as one can see from Fig.~\ref{v_eff_v_ref}.  Next, we fix $\tilde{\tau}_m=0.01$, and vary the threshold velocity.  As one can see in Fig.~\ref{p_ss}, there is always an abrupt change of slope in the vicinity of $\tilde{v}_{0}$.  For $\tilde{v}_{0} = 1$, the velocity distribution $p(\tilde{v})$ exhibits a two-peak structure, with a low and wide peak to the left of $\tilde{v}_{0}$, and a high and narrow peak to the right, and the average velocity of the swimmer is approximately $0.8 v_T$.  These results unequivocally demonstrate the feasibility of information swimming as a viable mechanism of self-propulsion.

In Fig.~\ref{v_eff_v_ref} we show how the average velocity $\langle \tilde v \rangle$ varies as a function of the threshold velocity $\tilde{v}_{0}$ for $\tilde{\tau}_m = 1, 0.1, 0.01$, and $\alpha^2  = 10, 2$ respectively.   In all cases we see that $\langle \tilde v \rangle $ vanishes as $\tilde v_0 \to \pm \infty$.  This is of course totally expected, since in these limits, measurement almost always return the same result, and the friction coefficient remains invariant.  The peaks of the curves in Fig.~\ref{v_eff_v_ref} correspond to the maximal average velocity achieved by tuning  $\tilde{v}_{0}$.  The location of the peak move to the right as $\alpha$ increases, or $\tilde \tau_m$ decreases.  However the optimal threshold velocity is never far away from zero.  The height of peak increases as $\alpha$ increases or $\tilde{\tau}_m$ decreases.  Note that the maximal velocity is generally less than, but of the same order as the thermal velocity.  

\vspace{2mm}
{\bf Entropic Efficiency of Swimming} \quad Similar to Maxwell's demon, an information swimmer record its measurement results, and as a consequence, the information entropy of its memory increases steadily during the motion.  Hence even though the entropy of the ambient fluid remains constant, the total entropy increases, in accordance with the second law.  Because information can be stored and transferred at arbitrary low temperature, increase of information entropy does not need to be accompanied by energy dissipation~\cite{landauer1961irreversibility}.  

We can quantify the rate of information entropy increase in swimmer's memory.  Let $s_n$ be the result of measurement at time $n \tau_m$, which takes $0$ if $v >v_{0}$ or $1$ if $v < v_{0}$.   The sequence $s_1, s_2, \ldots, s_n,\ldots$  forms a discrete Markov chain.  The results of consecutive measurements are however generally correlated, and hence the sequence can be compressed before storage.  According to information theory~\cite{cover2012elements,kullback1997information}, the minimal information bit needed to store each measurement result (averaged over long sequence of measurements) is the entropy rate of the Markov chain, which is defined as: 
\ba
I = - \sum p(s_{n+1},s_n)
\log_2\frac{p(s_{n+1},s_n)}{p(s_n)}. 
\label{info-rate-def}
\ea
The entropy production rate is then $\Sigma = I/\tau_m$, where $\tau_m$ is the period of measurement.  

Consider now a force-driven Brownian particle, whose dynamics is Eq.~(\ref{LE_original}) with $\alpha^2=1$, and augmented by an external force $F$:
$m\dot {v} = F - \gamma v + \sqrt {2 \gamma T} \, \zeta(t). $   The work done by the external force is constantly dissipated into the ambient fluid in the form of heat. The average velocity is $\langle v \rangle = F/\gamma$, and the entropy production rate is $\Sigma = F \langle v \rangle /T = \langle v \rangle^2 \gamma /T $.  Hence we have 
\be
{\langle v \rangle ^2}/ {\Sigma } = {T}/{\gamma}, \quad
 \mbox{Driven Brownian Particle}. 
\label{eta-0-def}
\ee 
 The same quantity for information swimmer is 
\be
 \frac{ \langle v \rangle ^2} {\Sigma }
 = \frac{\langle v \rangle ^2 \tau_m}{I }
 =  \frac {\langle \tilde v \rangle^2 \tilde{\tau}_m} { I} 
\cdot  \frac {T} {\gamma},  
\label{eta-1-def}
\quad \mbox{Information Swimmer.}
\ee

The ratio between Eqs.~(\ref{eta-1-def}) and (\ref{eta-0-def}) is:
\be
\eta_{IS} \equiv  \frac {\langle \tilde v \rangle^2 \tilde{\tau}_m} { I},
\label{eta-2-def}
\ee
which characterizes the entropic efficiency of information swimmer relative to a force driven Brownian particle with the same mass, baseline friction coefficient, and in the same temperature.  We compute this ratio for different values of control parameters.   As shown in Appendix B, the optimal choice of threshold velocity $\tilde{v}_0$ is always very close to zero.  Thus we fix $\tilde{v}_0=0$ to reduce to the task of computation.  In Fig.~\ref{swimmingEfficiency}, we plot ${\eta}_{\rm IS}$ as a function of  $\tilde{\tau}_m$ for $\alpha^2 = 2, 5, 10, 100$ respectively.  It is seen there that the optimal $\tau_m$ is  always a faction of $\tau$, and decreases monotonically as $\alpha$ increases.  The maximal efficiency monotonically increases with $\alpha$, and remains substantially lower than unity.  

\vspace{2mm}
{\bf  Information swimming on colored noise} \quad There are many systems where fluctuations exhibit long time-correlations.  For example, the time-correlations of velocity in fluids are characterized by long tails that decay algebraically~\cite{dorfman1994generic}.  Active fluids~\cite{ramaswamy2010mechanics,marchetti2013hydrodynamics} and turbulent fluids~\cite{mccomb1990physics} exhibit long range correlations both in time and in space.  These correlations can be used  to reduce the frequency of measurements for information swimmers. 

\begin{figure}[ht!]
	\centering
	\subfigure[]{\includegraphics[width=2.6in]{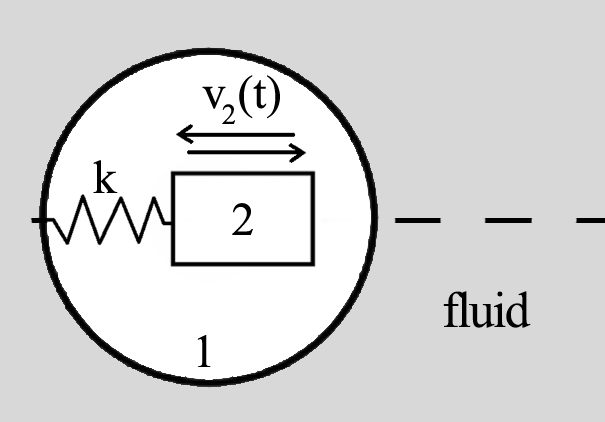}}
	\hspace{0mm}
	\subfigure[]{\includegraphics[width=2.6in]{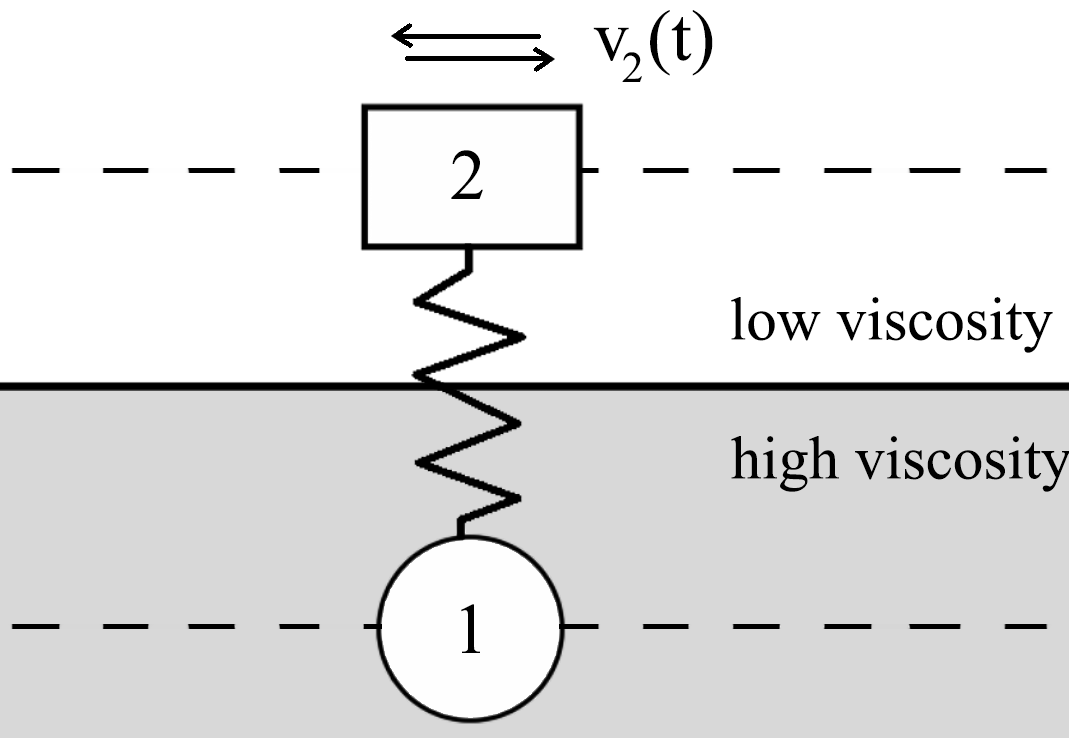}\label{2D_TE_alpha1over10}}
	\caption{Two toy models that realize the dynamics described by Eq.~(\ref{model-2-particles}).   Particles are constrained to move along the $x$ direction (dashed line) and are connected by a spring with spring constant $k$. 
	}
	\label{fig:coupled_two_particle_model}
\end{figure}

Here we study information swimmer using colored noises that are in thermal equilibrium.   As illustrated in Fig.~\ref{fig:coupled_two_particle_model}(a), we consider a system consisting of a box (with center-of-mass coordinate $x_1$ and negligible mass) and a particle (with coordinate $x_2$ and mass $m$) that are connected by a spring with constant $k$.  The particle is confined inside the box and therefore does not couple to noise or friction directly.  The dynamics of the two-body system is described by a set of coupled linear Langevin equations:
\begin{subequations}
 \ba
\gamma \, \dot{x}_1&=& -k(x_1-x_2 ) + \sqrt{ 2 \gamma T} \, \zeta(t),
\\
m \,\ddot{x}_2&=&- k \, (x_2-x_1). 
\ea
\label{model-2-particles}
\end{subequations}
where $\zeta(t)$ is a Gaussian white noise satisfying Eq.~(\ref{noise-variance-1}). Integrating out $x_1$, we find that the dynamics of $x_2$ satisfies the generalized Langevin equation with an effective {\em Ornstein-Uhlenbeck noise}:
\begin{subequations}
\ba
&& m \, \dot{v}_2(t) + \int_{0}^{\infty} \!\! K(  s ) {v_2}(t - s ) d s
= \eta(t).
 \label{GLE}\\
&& K(|t-s|) \equiv \frac{\gamma}{ \tau_{c}} 
\mathrm{e}^{-|t-s| / \tau_{c}},
\label{colored-noise}\\
 && \langle \eta(t)\rangle=0, \quad
\langle\eta(t) \eta(s)\rangle= T K(|t-s|),
\label{colored-noise-0}
\ea
\label{colored-noise-eff-model}
\end{subequations}
where $\tau_c = \gamma/k$ is the noise correlation time, which can be tuned continuously by tuning the spring constant $k$. In the limit $k \rightarrow \infty$, $\tau_c \rightarrow 0$, and Eq.~(\ref{colored-noise-eff-model}) reduces to the white noise model.  
 Details of the derivation are given by Appendix C. Equation (\ref{colored-noise-0}) is the {\em second Fluctuation-Dissipation Theorem} which relates variance of colored noise to the friction kernel \cite{kubo1966fluctuation}.  It is a consequence of the time-reversal symmetry of the original model Eq.~(\ref{model-2-particles}).    
 
Another possible realization of the dynamics (\ref{model-2-particles}) is illustrated in Fig.~\ref{fig:coupled_two_particle_model}(b), where two particles connected by a spring are moving near the interface of two fluids. The first particle moves in a fluid with high viscosity in the over-damped regime so that its mass can be ignored, and the second particle moves in a fluid with low viscosity so that both friction and noise can be ignored.

The velocity correlation function $\langle v_2(t)v_2(0) \rangle $ of the particle can be calculated using standard Laplace transform or Fourier transform (See Appendix C).  For $\tilde k \equiv m k/\gamma^2 = \tau/\tau_c < 4 $,  $\langle v_2(t)v_2(0) \rangle $ exhibit oscillation \cite{hansen2013theory}:
\ba
\langle v_2(t)v_2(0) \rangle = \frac{T}{m}
e^{-\frac{kt}{2\gamma}}\left(\cos(\Omega t)+\frac{k/\gamma}{2\Omega}\sin(\Omega t)\right),
\ea
where $\Omega \equiv \sqrt{k/m-k^2/(4\gamma^2)}$.

\begin{figure}
	\centering
	\includegraphics[width=12.5cm]{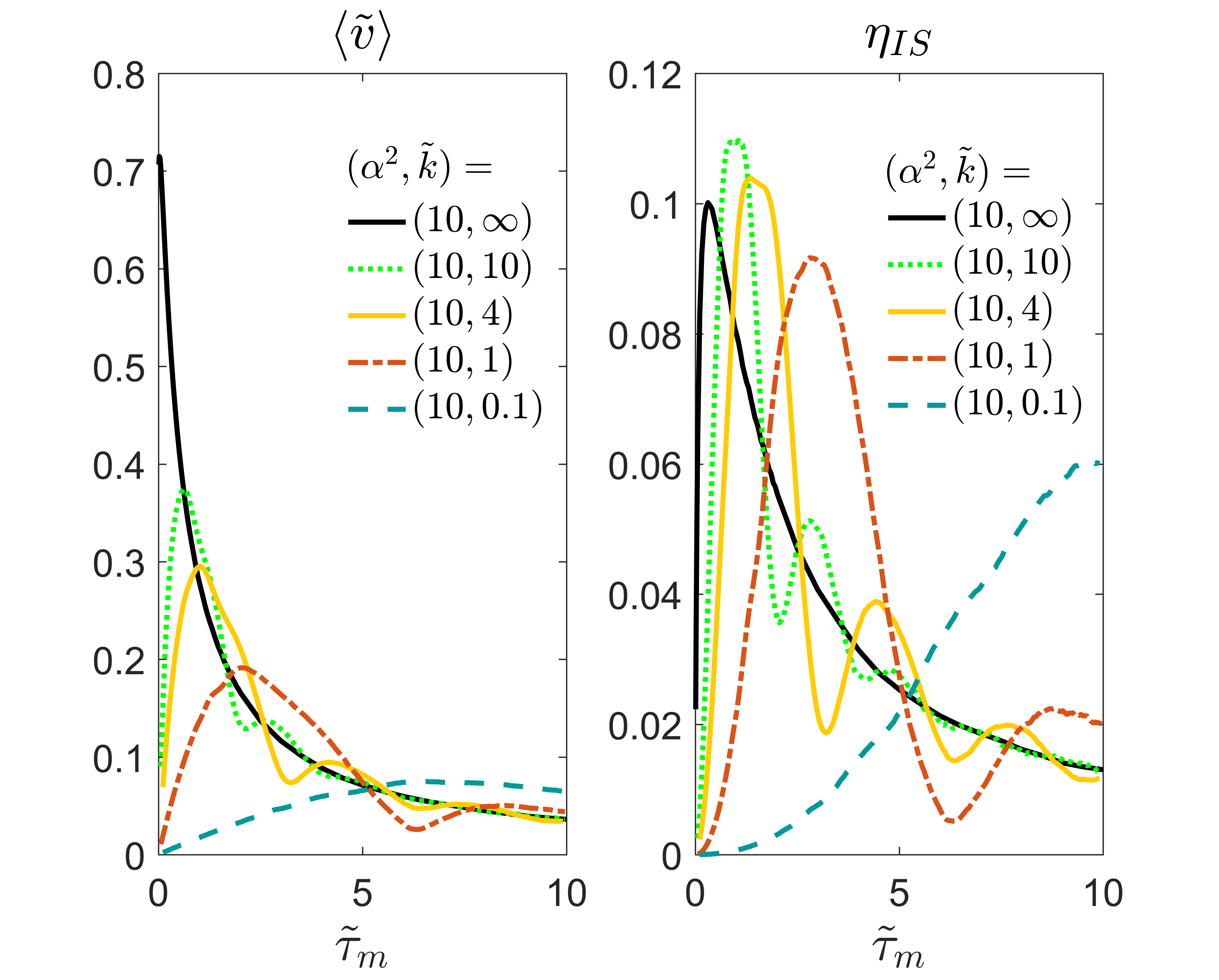}
	\caption{(Color online) The average velocity $\langle\tilde{v}\rangle$ (left panel) and entropic efficiency $\eta_{IS}$ (right panel). We fix $\tilde{v}_0=0$ and run simulation for $\alpha^2=10$ with various $\tilde{k} = \tau/\tau_c$. The line with $\tilde{k}\to\infty$ corresponds to the white noise.}
	\label{eta_swim_colored}
\vspace{-4mm}
\label{fig:colored-noise}
\end{figure}

We can now introduce a measure-feedback mechanism into the model Eqs.~(\ref{model-2-particles}), so that it becomes an information swimmer. The system measures velocity $v_2$ every $\tau_m$ second, and tunes the friction coefficient to $\gamma$ if $v_2 > v_0$ and $\alpha^2 \gamma$ if $v_2 < v_0$.  
The coupled Langevin equations  (\ref{model-2-particles}) are simulated using the same method as above.  The dimensionless variables are defined the same as in Eq.~(\ref{rescaling-1}).  
In Fig.~\ref{fig:colored-noise}, we show respectively the average velocity $\langle\tilde{v}\rangle$ and entropic efficiency $\eta_{IS}$ as a function of period of measurement $\tilde \tau_m$ for various values of $\tilde k$.  It is seen that as long as $\tilde k$ is finite, both quantities exhibit oscillation as a function of $\tilde \tau_m$.   These oscillations can be used to design information swimmers that optimize velocity or efficiency.  Furthermore, it appears that both quantities converge to zero as $\tilde \tau_m \rightarrow 0$, indicating that the measurement-feedback mechanism becomes ineffective as the frequency of measurements becomes high, whose reason we do not yet understand.   Finally, as the correlation time of noise becomes longer and longer (with decreasing $\tilde k$), the peaks of curves move systematically towards the right in the $\tilde \tau_m$ axis.  We note that while the maximal velocity decreases steadily as $\tau_c$ increases, the change of maximal efficiency is very insignificant.  The general conclusion is therefore we can use correlation of noises to reduce the frequency of measurement without changing the swimming efficiency.  This may be very useful for design of information swimmers in non-equilibrium environment, such as turbulent fluids~\cite{mccomb1990physics} or active fluids~\cite{ramaswamy2010mechanics,marchetti2013hydrodynamics}.

X.X. acknowledge support from NSFC via grant \#11674217, as well as additional support from a Shanghai Talent Program.  This research is also supported by Shanghai Municipal Science and Technology Major Project (Grant No.2019SHZDZX01).

\appendix

\section{Numerical integrator}
The Langevin equation with the Gaussian white noise $\tilde{\zeta}(\tilde{t})$ satisfying $\langle  \tilde \zeta(\tilde t) \rangle = 0,\ 
\langle \tilde \zeta(\tilde t) \tilde \zeta(\tilde t') \rangle 
= 2 \, \delta(\tilde t- \tilde t')$ is
\ba
\frac{d \tilde{v}(\tilde{t})}{d \tilde{t}} = \tilde{F}(\tilde{x}(\tilde{t})) - \beta^2 \tilde{v}(\tilde{t}) + \beta \, \tilde{\zeta}(\tilde t)
\ea
where $\tilde{F}(\tilde{x}(\tilde{t}))$ is a position-dependent force. We discretize the equation over the small time step $\Delta \tilde{t}$. The value of $\beta\in\{1,\alpha\}$ is determined from measurement results of velocity $\tilde{v}$ every $l$ time steps when $l\Delta\tilde{t}=\tilde{\tau}_m$, and is kept fixed during those $l$ time steps. We can integrate the equations using the first-order Euler-Maruyama scheme \cite{kloeden2013numerical}: \ba \tilde{x}(\tilde{t}+\Delta \tilde{t})&=&\tilde{x}(\tilde{t})+\Delta \tilde{t}\tilde{v}(\tilde{t}),\nonumber\\
\tilde{v}(\tilde{t}+\Delta \tilde{t})&=&\tilde{v}(\tilde{t})+\Delta \tilde{t}[\tilde{F}(\tilde{x}(\tilde{t}))-\beta^2\tilde{v}(\tilde{t})]+\beta W\sqrt{2\Delta \tilde{t}}\label{numerical}\ea 
where $W$ is a Gaussian variable with average zero and standard deviation one. There are also second-order integrators, such as the one-step collocation via the Taylor expansion \cite{mannella1989fast}:
\ba
\tilde{x}(\tilde{t}+\Delta \tilde{t}) &=&\tilde{x}(\tilde{t})+\Delta \tilde{t} \tilde{v}(\tilde{t})+C(\tilde{t}) \nonumber\\ \tilde{v}(\tilde{t}+\Delta \tilde{t}) &=&\tilde{v}(\tilde{t}) +\Delta \tilde{t} [\tilde{F}(\tilde{x}(\tilde{t}))-\beta^2 \tilde{v}(\tilde{t})]+\beta W\sqrt{2\Delta \tilde{t}} -\beta^2 C(\tilde{t}) \nonumber\\ C(\tilde{t}) &=&\frac{\Delta \tilde{t}^{2}}{2}[\tilde{F}(\tilde{x}(\tilde{t}))-\beta^2 \tilde{v}(\tilde{t})]+\beta\sqrt{2} \Delta \tilde{t}^{3 / 2}\left(\frac{1}{2} W+\frac{1}{2 \sqrt{3}} V\right)
\ea
where $W,V$ are two uncorrelated Gaussian variables with average zero and standard deviation one. A simple comparison of the two numerical schemes shows that they give the same results in our main text, so we just use the Euler-Maruyama scheme in the simulation. Also note that the numerical integrator for the overdamped dynamics can be similarly obtained. The time step is set to $\Delta \tilde{t}=0.001$ throughout the simulation, which is neither too large to make the numerical method invalid nor too small to slow down the simulation.

\section{Entropic efficiency, entropy rate and effective velocity}
\begin{figure}[ht!]
	\centering
	\subfigure[]{\includegraphics[width=2.6in]{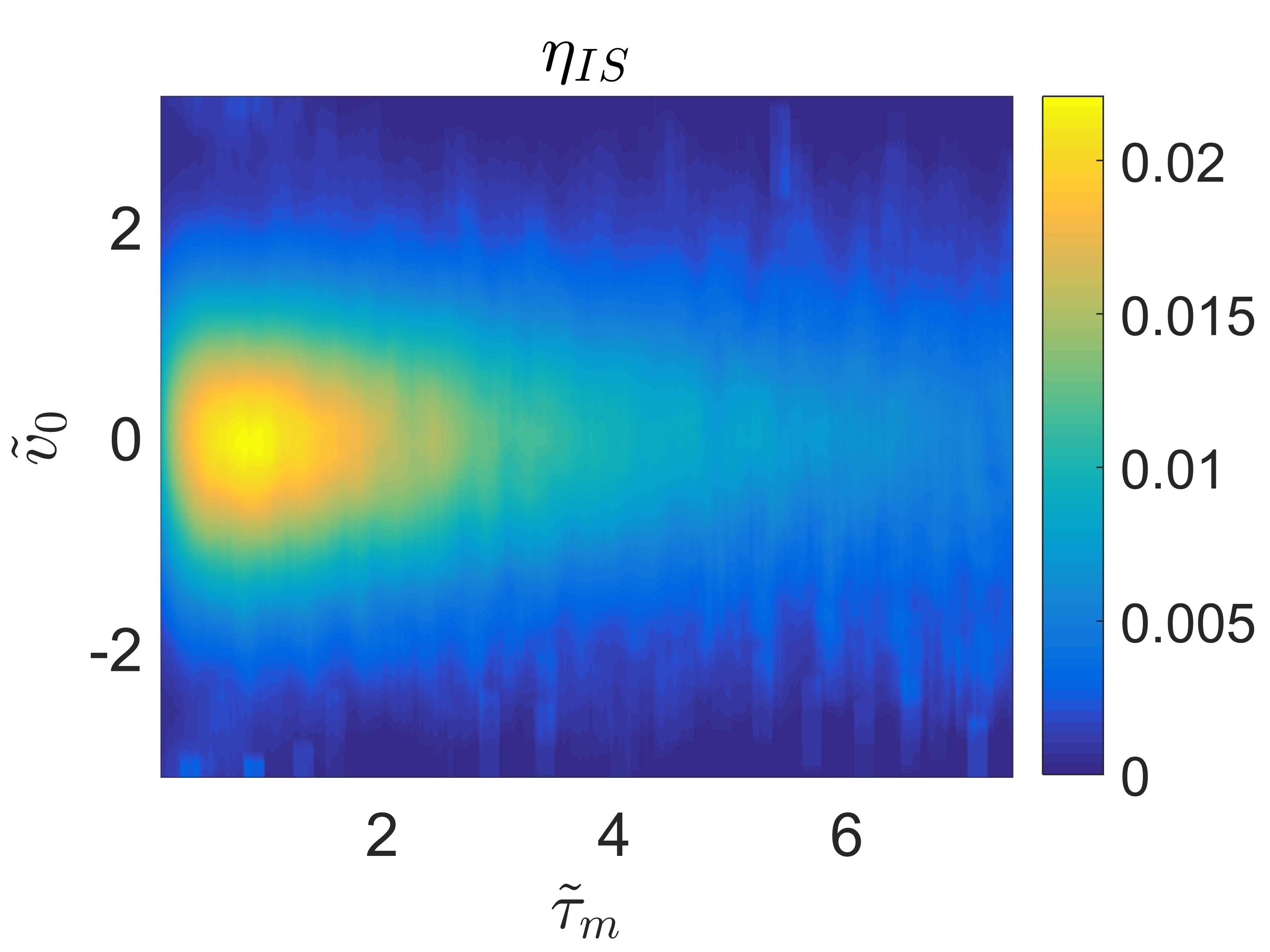}\label{2D_eta_alpha0_5}}
	\hspace{-3.8mm}
	\subfigure[]{\includegraphics[width=2.6in]{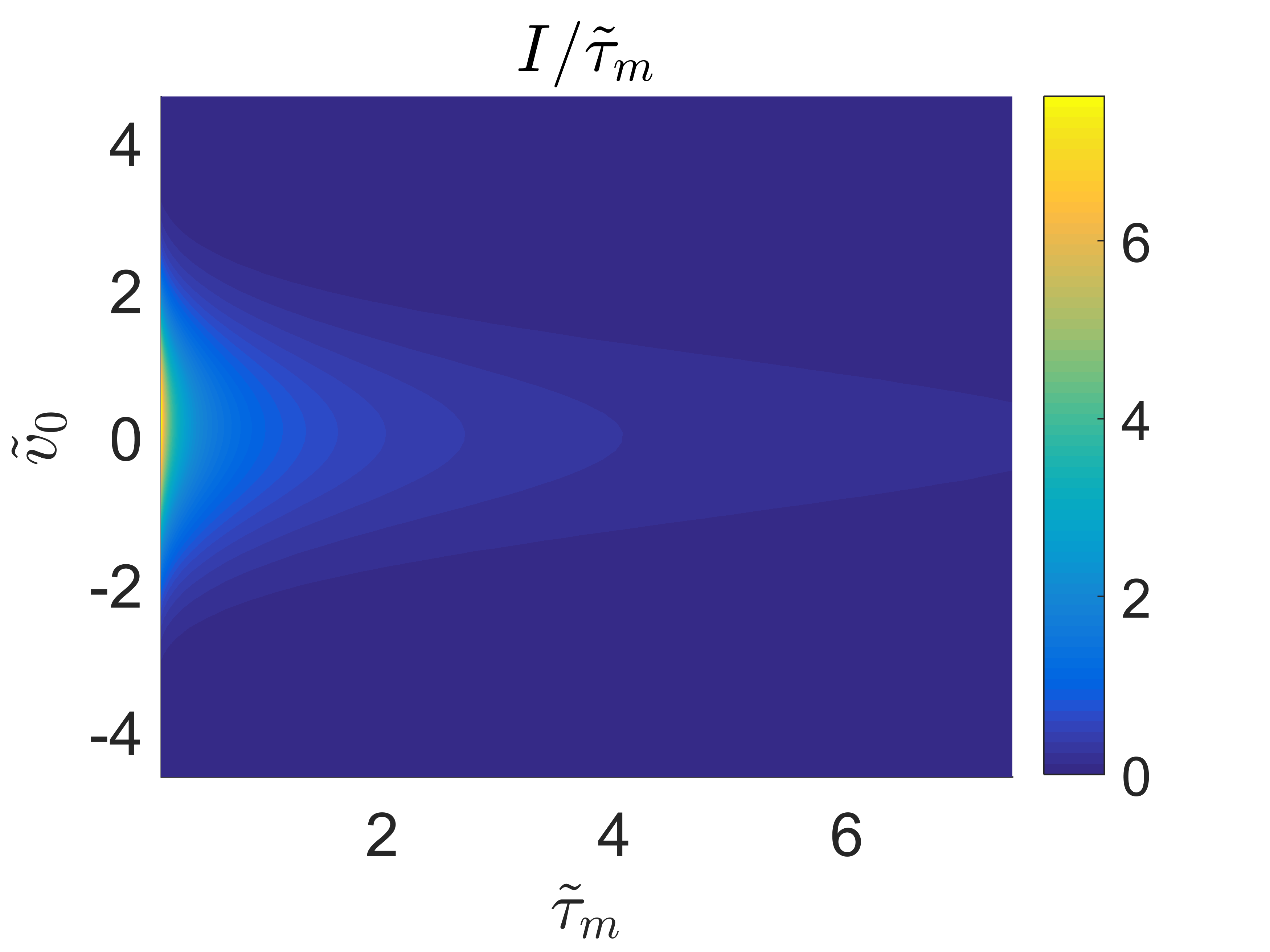}\label{2D_TE_alpha1over10}}
	\hspace{-3.8mm}
	\subfigure[]{\includegraphics[width=2.6in]{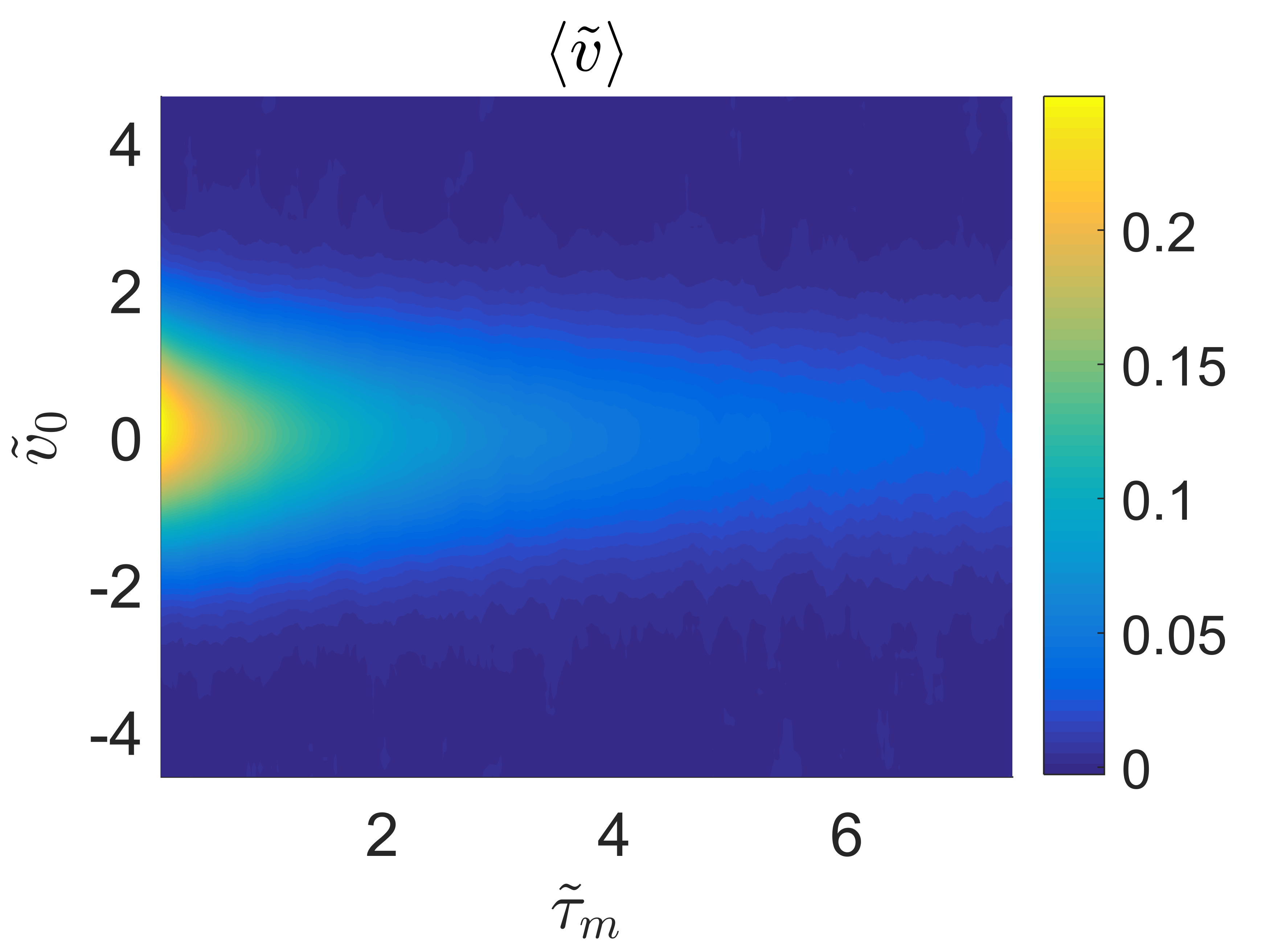}\label{2D_v_eff_alpha1over10}}
	\caption{(Color online) The ratio $\eta_{IS}$, the value $I/\tilde{\tau}_m$ and effective velocity $\langle \tilde v \rangle$ as functions of $\tilde{v}_{0}$ and $\tilde{\tau}_m$.}
	\label{2D}
\end{figure}

In Fig.~\ref{2D} we plot the entropic efficiency $\eta_{IS}$, the rescaled entropy rate $I/\tilde{\tau}_m$ and the effective velocity $\langle \tilde{v} \rangle$ as functions of both $\tilde{v}_{0}$ and $\tilde{\tau}_m$, with $\alpha^2=2$. The efficiency is always less than unity, and achieves maximum at $\tilde \tau_m \approx 1$. It is shown that the optimal choice of threshold velocity $\tilde{v}_0$, which maximize the entropic efficiency $\eta_{IS}$, is always very close to zero, thus we fix $\tilde{v}_0=0$ without loss of generality.

\section{Information Swimmer with Colored Noises}

In this appendix, we present some details about the two-particle toy model discussed in the main text. The dynamics is described by the following set of linear Langevin equations
\begin{subequations}
	\label{LinearLD_x1}
	\ba	
	\gamma\dot{x}_1&=& -k(x_1-x_2) + \sqrt{2\gamma T} \zeta(t),
	\label{LinearLD_x1-1}
	\\
	m\ddot{x}_2 &=& - k (x_2-x_1)
	\label{LinearLD_x1-2}
	\ea
	where $\zeta(t)$ is the normalized Gaussian white noises satisfying $\langle \zeta(t)\rangle = 0$, $\langle \zeta(t) \zeta(t') \rangle =  \delta(t - t')$.
\end{subequations}
These equations can be Fourier transformed into
\ba
- i \gamma \omega x_1(\omega) &=&
- k (x_1(\omega) - x_2(\omega))
+ \sqrt{2\gamma T} \zeta(\omega),
\label{LinearLD_x1-1-omega}
\\
- m \omega^2 {x}_2(\omega) &=& - k (x_2(\omega) -x_1(\omega) ).  
\label{LinearLD_x1-2-omega}
\ea

We are only interested in the dynamics of particle 2.  For this purpose, we use Eq.~(\ref{LinearLD_x1-1-omega}) to express $x_1(\omega)$ in terms of $x_2(\omega)$, and substitute it back into Eq.~(\ref{LinearLD_x1-2-omega}) so that we obtain an equation for $x_1(\omega)$:
\ba
\left( - m \omega^2 + \frac{- i k \omega}{- i \omega + k/\gamma} \right)
x_2(\omega) = \frac{k}{\gamma} \cdot
\frac{\sqrt{2 \gamma T} }{ - i \omega + k/\gamma}
\zeta(\omega). 
\ea
But this can be further written as an equation for $v_2(\omega) = - i \omega x_2(\omega)$:
\ba
\left( - i m \omega + \frac{k}{- i \omega + k/\gamma} \right)
v_2 (\omega) = \frac{k}{\gamma} \cdot
\frac{\sqrt{2 \gamma T} }{ - i \omega + k/\gamma}
\zeta(\omega).  
\label{LinearLD_x1-3-omega}
\ea

We can further define an effective colored noise:
\ba
\eta (\omega ) &=& \frac{k}{\gamma} \cdot
\frac{\sqrt{2 \gamma T} }{ - i \omega + k/\gamma}
\zeta(\omega). 
\ea
whose power spectrum is given by:
\ba
S_{\eta}(\omega) = \frac{2 k^2 T/\gamma}{\omega^2 + k^2/\gamma^2}. 
\ea
The noise correlation function in time domain is then given by the Fourier transform of  $S_{\eta}(\omega) $:
\ba
\langle \eta(t) \eta(t') \rangle = \int \frac{d \omega}{2 \pi}
S_{\eta}(\omega) \, e^{- i \omega (t - t') }
= k T\, e^{- k |t|/\gamma}.
\label{eta-correlation-1}
\ea
Finally Eq.~(\ref{eta-correlation-1}) shows that $\eta(t)$ is the {\em Ornstein-Uhlenbeck} colored noise. 

Let us now define $\tau_c = \gamma/k$, and a friction kernel $K(t)$:
\ba
K(t) &=& \left\{ \begin{array} {ll}
	\frac{\gamma}{\tau_c}\, e^{- t/\tau_c}, & t \geq 0
	\vspace{3mm}\\
	0, & t < 0
\end{array}
\right.
\label{K-correlation-def}
\ea
which has the Fourier transform:
\be
K(\omega) = \frac{k \gamma}{- i \gamma \omega + k}
= \frac{\gamma/\tau_c}{- i \omega + \tau_c^{-1}},
\ee
which has the following symmetry:
\ba
K(- \omega) = K(\omega)^*.
\ea
We can easily show that this kernel $K(\omega)$ is related to the noise spectrum $S_\eta(\omega) $ via 
\ba
S_\eta(\omega)  = 2 T\, {\rm Re}\, K(\omega). 
\ea
We can rewrite Equations (\ref{LinearLD_x1-3-omega}) and (\ref{eta-correlation-1}) in the following forms:
\ba
&& m \dot v_2(t) + \int_0^\infty dt' K(t' ) v_2(t-t') = \eta(t), 
\label{LinearLD_x1-4-omega}
\\
&& \langle \eta(t) \eta(t') \rangle = T\, K(|t - t'|)
= T\,( K(t-t') +  K(t'-t) ). 
\label{correlation-eta-1}
\ea
Equation (\ref{LinearLD_x1-4-omega}) shows that the dynamics of $x_2(t)$ can be described by a generalized Langevin theory with an effective noise $\eta(t)$ which has {\em Ornstein-Uhlenbeck}  spectrum.  Equation (\ref{correlation-eta-1}) is the second fluctuation-dissipation theorem which relates noise correlation to the friction kernel.   It is a consequence of the time-reversal symmetry of the original two-particle model (\ref{LinearLD_x1}).  

Let us note that Eq.~(\ref{LinearLD_x1-3-omega}) can also be written as 
\be
\left( -\frac{m \gamma}{k} \omega^2 
- i m \omega + \gamma \right) v_2 (\omega)
=  \sqrt{2 \gamma T} \zeta(\omega).
\label{dynamics-v2-Fourier}
\ee
Making Fourier transform we further obtain
\ba
\frac{m \gamma}{k} \ddot v_2(t) + m \dot v_2(t) + \gamma v_2(t)
= \sqrt{2 \gamma T} \zeta(t).
\label{dynamics-v2}
\ea

For simplicity, we introduce dimensionless variables. Set the time scale $\tau=m/\gamma$ and the velocity scale $v_T=\sqrt{T/m}$, the dimensionless variables are
\ba
\tilde{t}=\frac{t}{\tau},\quad 
\tilde{v}(\tilde{t}) = \frac{v(t)}{v_T},\quad 
\tilde{\zeta}(\tilde{t})=\sqrt{\frac{2m}{\gamma}}\zeta(t),\quad
\tilde{k}=\frac{m}{\gamma^2}k
\ea
Eq.~(\ref{dynamics-v2}) can be rewritten as:
\ba
(- \tilde{\omega}^2 - i \tilde{\omega} \tilde{k} 
+ \tilde{k})\hat{\tilde{v}}_2(\tilde{\omega}) 
= \tilde{k} \hat{ \tilde{ \zeta} } (\tilde{\omega})
\ea
where $\tilde{\omega}$ is the dimensionless frequency, reciprocal to the dimensionless time $\tilde t$. We can solve for $\tilde{v}_2$ in terms of the noise to obtain
\ba
\hat{\tilde{v}}_2(\tilde{\omega})=\frac{\tilde{k}\hat{\tilde{\zeta}}(\tilde{\omega})}{-\tilde{\omega}^2-i\tilde{\omega} \tilde{k}+\tilde{k}}
\ea
The spectral density of $\hat{\tilde{\zeta}}$ is the Fourier transform of its correlation function
\ba
S_{\tilde{\zeta}}(\tilde{\omega})=\int_{-\infty}^{\infty}d\tilde{t} e^{i\tilde{\omega} \tilde{t}}\langle\tilde{\zeta}(\tilde{t})\tilde{\zeta}(0)\rangle=2
\ea
The spectral density of $\tilde{v}_2$ is proportional to
\ba
S_{{\tilde{v}_2}}(\tilde{\omega})=\frac{\tilde{k}^2S_{\tilde{\zeta}}(\tilde{\omega})}{(\tilde{k}-\tilde{\omega}^2)^2+\tilde{\omega}^2 \tilde{k}^2}
\ea
The time dependent correlation function is
\ba
C_{\tilde{v}_2}(\tilde{t}) = 
\frac{1}{2\pi}\int_{-\infty}^\infty d\tilde{\omega} 
e^{-i\tilde{\omega} \tilde{t}}S_{\tilde{v}_2}(\tilde{\omega})
=\frac{\tilde{k}^2}{\pi}\int_{-\infty}^\infty d\tilde{\omega} 
e^{-i\tilde{\omega} \tilde{t}}\frac{1}
{(\tilde{k} -\tilde{\omega}^2)^2+\tilde{\omega}^2 \tilde{k}^2}.
\label{contour-integral-1}
\ea
The integrand has four poles in the complex plane:
\ba
\tilde{\omega}=\pm \frac{ i }{2} \,\tilde{k} \pm \tilde{\omega}_1, 
\quad
\tilde{\omega}_1 \equiv \sqrt{\tilde{k}-\tilde{k}^2/4}.  
\ea

We only need to calculate the correlation function $C_{\tilde{v}_2}(\tilde{t})$ for $\tilde{t} > 0$. The contour integral in Eq.~(\ref{contour-integral-1}) then has to be closed in the lower half plane.  The resulting correlation function may decay monotonically or oscillatorily, depending on the value of $\tilde k$.  
\begin{itemize}
	\item If $ \tilde{k} = mk/\gamma^2 >4$, all poles are purely imaginary, and the correlation function decays monotonically: 
	\ba
	C_{\tilde{v}_2}(\tilde{t})&=&
	\frac{\tilde{k} \, e^{-\tilde{t}(\tilde{k}/2-|\tilde{\omega}_1|)}  }
	{|\tilde{\omega}_1|(\tilde{k}-2|\tilde{\omega}_1|)} 
	- \frac{ \tilde{k} \, e^{-\tilde{t}(\tilde{k}/2 + |\tilde{\omega}_1|)}} 
	{|\tilde{\omega}_1|(\tilde{k}+2|\tilde{\omega}_1|)} . 
	\label{exponential-2-model}
	\ea
	
	\item  If $0 < \tilde{k} = mk/\gamma^2 < 4$, the poles are not on the imaginary axis, and the   $v_2$ correlation function is oscillatory:
	\ba
	C_{\tilde{v}_2}(\tilde{t})
	&=& e^{-\tilde{k}\tilde{t}/2}
	\left(\cos(\tilde{\omega}_1\tilde{t})
	+\frac{\tilde{k}}{2\tilde{\omega}_1}
	\sin(\tilde{\omega}_1\tilde{t})
	\label{oscillatory-2-model} \right). 
	\ea
	
	\item If $\tilde{k}= mk/\gamma^2 = 4$, the poles become degenerate, and the correlation function becomes
	\ba
	C_{\tilde{v}_2}(\tilde{t})=e^{-2\tilde{t}}(1+2\tilde{t})=e^{-2{t}/\tau}(1+2{t}/\tau)
	\ea
\end{itemize}

\begin{figure}[ht!]
	\centering
	\subfigure[]{\includegraphics[width=3.3in]{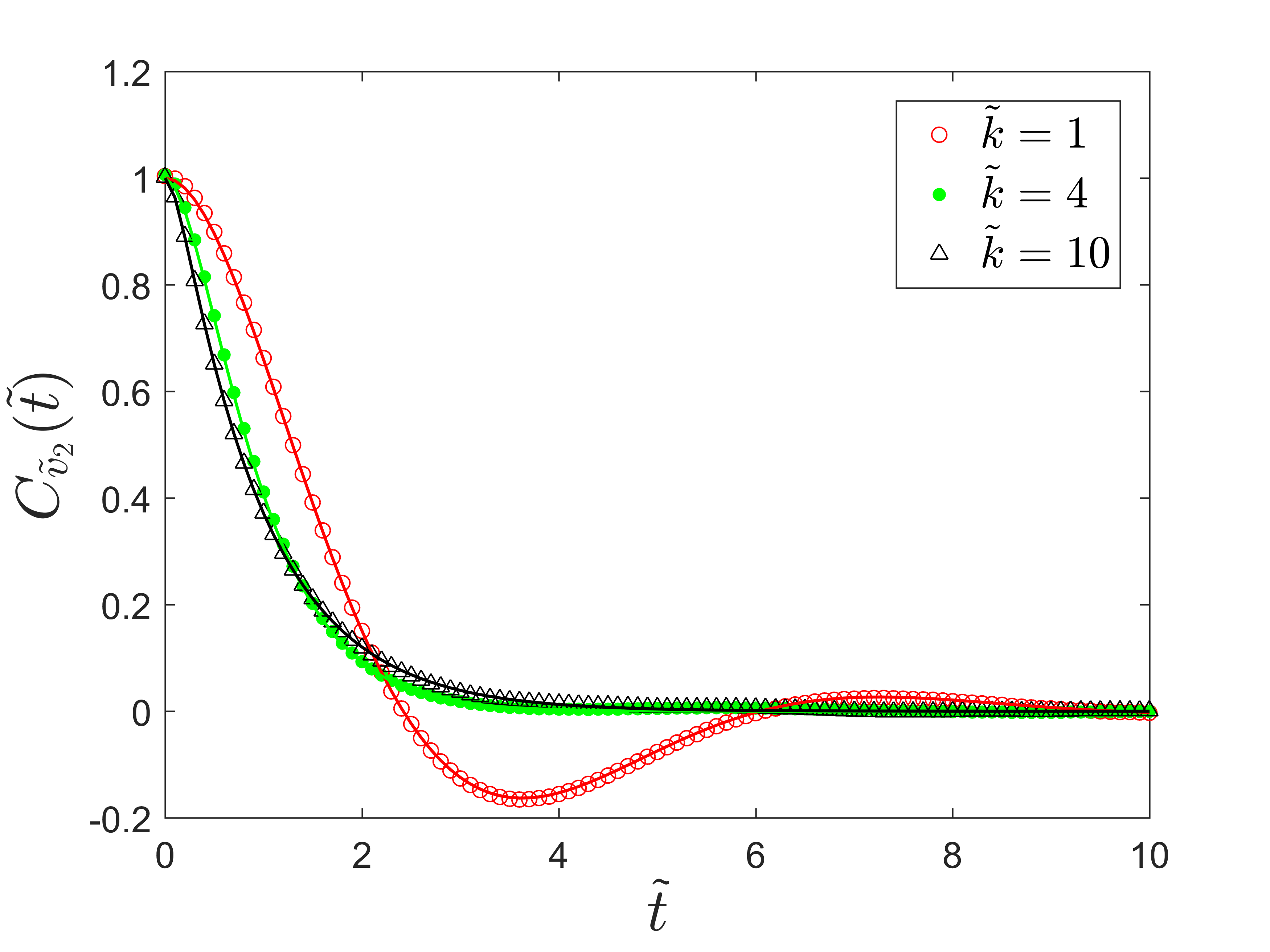}\label{2D_eta_alpha0_5}}
	\hspace{-3.8mm}
	\subfigure[]{\includegraphics[width=3.3in]{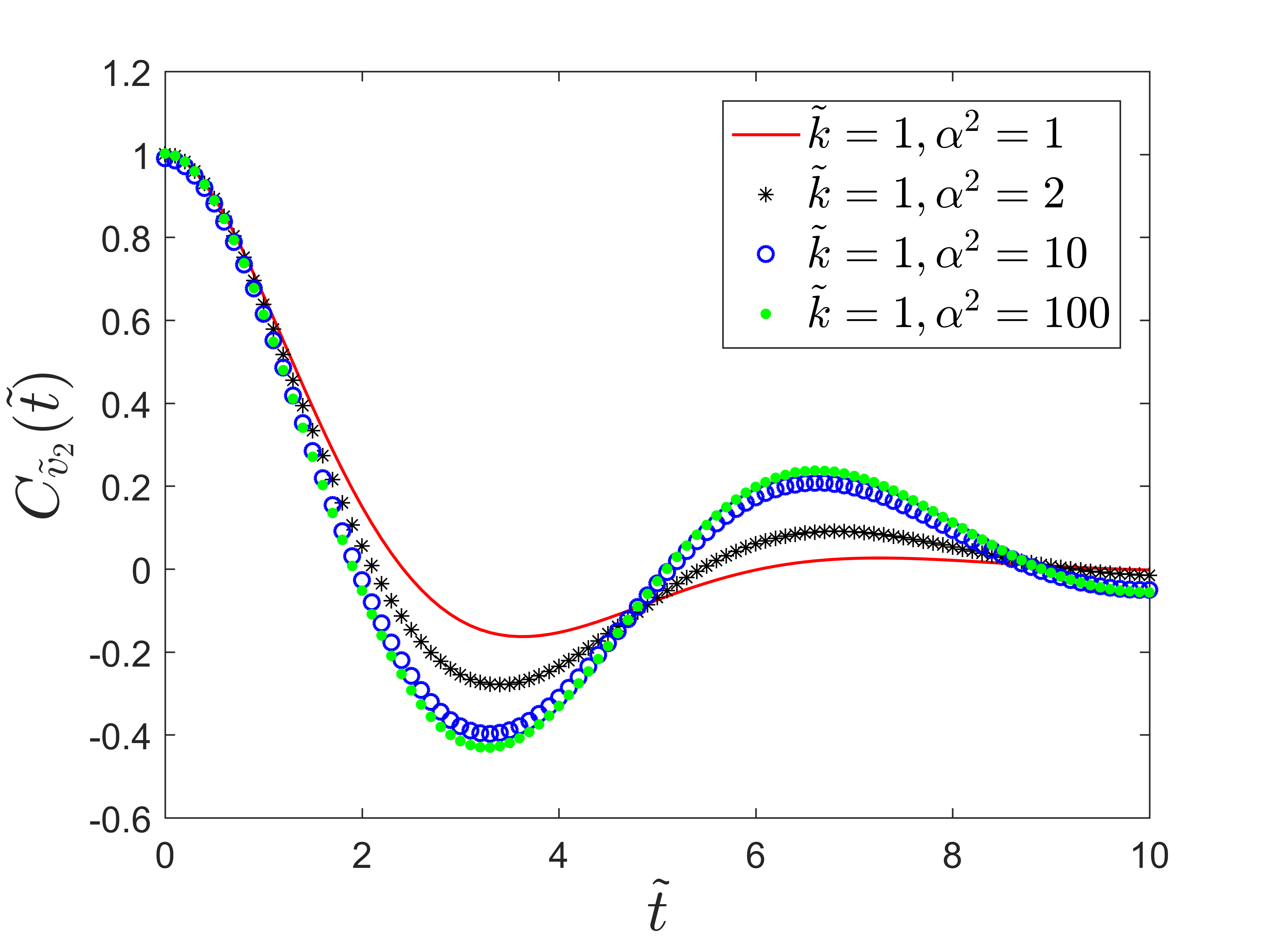}\label{2D_TE_alpha1over10}}
	\caption{(Color online) (a) Velocity correlation function $C_{\tilde{v}_2}(\tilde{t})$ for the two-particle model Eq.~(\ref{LinearLD_x1}) in equilibrium with $\alpha^2=1$.  The symbols are numerical simulation results, whereas the solid lines are theoretical curves. (b) Velocity correlation functions for information swimmer, Eq.~(\ref{Info-swimmer-colored}), with $\tilde{k}=1$. The red line is the equilibrium theoretical curve with $\tilde{k}=1, \alpha^2=1$.}
	\label{v2_corr}
\end{figure}

We can now introduce the information swimming mechanism into the two-particle toy model.  The system measures velocity $v_2$ every $\tau_m$ second, and tunes the friction coefficient $\beta^2\gamma$ such that $\beta\in\{1,\alpha\}$ according to the result of measurement.  In the time interval $n \tau_m < t < (n+1)\tau_m$, Eq.~(\ref{LinearLD_x1}) becomes
\begin{subequations}
	\label{Info-swimmer-colored}
	\ba	
	\beta^2\gamma\dot{x}_1&=&
	- k(x_1-x_2)+\beta\sqrt{2\gamma T}\zeta(t)\\
	m\ddot{x}_2&=&-k(x_2-x_1)
	\ea
	where the parameter $\beta $ is set to be 
	\ba
	\beta=\begin{cases}
		\alpha & \text{if } \,\,\, v_2 (n \tau_m) <v_0,  
		\vspace{3mm}\\
		1 & \text{if } \,\,\, v_2 (n \tau_m)  \geq v_0.
	\end{cases}
	\ea
\end{subequations}
during this interval, and we use $v_0=0$ throughout the simulation. In Fig.~\ref{v2_corr} we compare the correlation functions for the dimensionless variables $C_{\tilde{v}_2}(\tilde{t})$ from simulation with theory, both in the absence ($\alpha^2=1$) and presence ($\alpha^2\neq1$) of measurement. The correlation functions of the equilibrium state ($\alpha^2=1$) from simulation match their theoretical counterparts as shown in Fig.~\ref{v2_corr}(a). In Fig.~\ref{v2_corr}(b) with information swimming mechanism and at $\tilde{k}=1$, if we increase $\alpha^2$ from 2 to 100, the velocity correlation would vibrate at a larger and larger magnitude. 

\end{document}